\documentstyle[12pt]{article}

\makeatletter
\def\@begintheorem#1#2{\it \trivlist \item[\hskip \labelsep{\bf #1\ #2.}]}
\newtheorem{teo}{Theorem}[section]

\newtheorem{prop}[teo]{Proposition}

\newfont{\Bbb}{msbm10 scaled 1200}
\def\mr{{\hbox{\Bbb R}}}

\def\mh{{\hbox{\Bbb H}}}

\def\mn{{\hbox{\Bbb N}}}

\def\mm{{\hbox{\Bbb M}}}

\def\mx {{\hbox{\Bbb X}}}

\newfont{\Got}{eufm10 scaled 1200}

\title{Geometric Cone Surfaces and (2+1) - Gravity coupled to Particles}

\author{Riccardo Benedetti and Enore Guadagnini}

\begin{document}

\maketitle

\begin{abstract} 
\noindent We introduce the (2+1)-spacetimes with compact space of
genus $g\geq 0$ and $r$ gravitating particles which arise by {\it
Minkowskian suspensions} of flat or hyperbolic cone surfaces, by {\it
distinguished deformations} of hyperbolic suspensions and by {\it
patchworking} of suspensions. Similarly to the matter-free case, these
spacetimes have nice properties with respect to the canonical
Cosmological Time Function. When the values of the masses are
sufficiently large and the cone points are suitably spaced, 
the distinguished deformations of
hyperbolic suspensions determine a relevant open subset of the full
parameter space; this open subset is  homeomorphic to 
${\cal U}\times \mr^{6g-6+2r}$, where
$\cal U$ is a non empty open set of the Teichm\"uller Space $T^r_g$.
By patchworking of suspensions one can produce examples of spacetimes
which are not distinguished deformations of any hyperbolic
suspensions, although they have the same masses; in fact, we will
guess that they belong to different connected components of the
parameter space.
\end{abstract}

\section{Introduction.}

\noindent Globally hyperbolic matter-free (2+1) - spacetimes with
compact space of genus $g\geq 1$ and cosmological constant $\Lambda =
0$ have been fairly well understood. These spacetimes can be arranged
into classes up to {\it Teichm\"uller equivalence} that is, roughly
speaking, up to isometry isotopic to the identity. Both the {\it
geometric-time-free} approach, which eventually identifies each
spacetime by its {\it geometric holonomy} (c.f. {\bf [W], [Me]}), 
and the {\it cosmological} approach based on
fibration by {\it constant mean curvature} space-like surfaces (see
{\bf [Mo], [A-T-M]}), have recognized the {\it Cotangent Bundle of
the Teichm\"uller space $T_g$} (which is homeomorphic to
$B^{6g-6}\times \mr^{6g-6}$, when $g\geq 2$) as the
parameter space. When $g\geq 2$, the correspondence between these two
approaches is still rather implicit; nevertheless, we have shown in
{\bf [B-G 1]} that the canonical {\it Cosmological Time Function}
(CTF) (that is the {\it length of time that the events have been in
existence}) provides a very good cosmological resolution of the
matter-free (2+1) gravity. For instance, the {\it asymptotic states}
of (CTF) recover and decouple the linear part and the translation part
of the geometric holonomy; the orbit in $T_g$ of the (CTF) has nice
properties; the ``initial'' singularity with respect to (CTF) can be
accurately described in terms of the degeneration of the geometry of the
level surfaces.

\noindent 
It turns out that all matter-free spacetimes are obtained by means of
two basic constructions: (a) the {\it Minkowskian suspension} of flat or
hyperbolic surfaces; (b) a distinguished kind of deformations of the
hyperbolic suspensions. In this paper we shall describe the extension of
constructions (a) and (b) to the case of gravitating particles in (2+1)
dimensions.

\medskip

\noindent 
When gravity is coupled to particles, the picture is far from being
exhaustive. We will be concerned with {\it compact spaces} with a
finite number of massive particles. 't Hooft's approach {\bf ['t H]}
describes these spacetimes by means of the ``linear'' evolution of a
special kind of Cauchy surfaces which are tiled by spatial planar
polygons. The extrinsic curvature is null in the interior of each tile
and it is singular along the edges; the evolution includes the
changing of tiling combinatorics under codified transition rules. Each
such a Cauchy surface is intrinsecally a flat surface with conical
singularities. Among these singularities, some correspond to the
intersection with particle world-lines (the spacetime has a
concentrated curvature along these lines); the others are
3-dimensionally {\it apparent singularities} (but the Gauss-Bonnet
constraint implies that, in general, they cannot be avoided ).  Each
globally hyperbolic spacetimes contains such a kind of Cauchy surface
with, at least locally, such a kind of evolution; it is not clear to
us if the evolution of a given surface necessarily fills all the
spacetime and how the evolutions of different surfaces in the same
spacetime are related each other. So, it seems hard to recover from
this approach a clear identification of the parameter space.

\noindent Another experimented approach (see {\bf [Mn-S], [B-C-V]}) 
is the classical ADM formalism with the so called
``instantaneaous gauge'', that requires fibration by spatial Cauchy
surface with zero extrinsic curvature. This last requirement is
technically very useful and allows to analytically find solutions by means
of classical and very elegant mathematical tools. Unfortunately, it
turns out that the only spacetimes with compact space covered by this
approach are the static ones (that is, by using the terminology of the
present paper, the static Minkowskian suspensions of flat surfaces
with conical singularities, which we will see below).
\medskip

\noindent The aim of this note is to describe the spacetimes with
compact space of genus $g\geq 0$ and $r$ gravitating particles
($\Lambda = 0$) that one can obtain by means of three kinds
of construction: (a) the {\it Minkowskian suspensions of flat or
hyperbolic surfaces with conical singularities}; (b) the {\it
distinguished deformations} of hyperbolic suspensions (in strict
analogy with the matter-free case); (c) the {\it patchworking} of
Minkowskian suspensions (this is peculiar to gravity coupled with
particles).  These spacetimes have very transparent structural
properties and behave somewhat similarly to the matter-free ones with
respect to, for instance, the canonical (CTF), its asymptotic states,
the initial singularity and so on.

\noindent Moreover they form a rather wide class of spacetimes, so
that we can derive from them some non trivial information about the
actual parameter space. For example we will show that when the masses
are big enough and the cone points are suitaby spaced (roughly
speaking), then the distinguished deformations of hyperbolic
suspensions determine a relevant non empty open subset of the
parameter space of the form ${\cal U}\times \mr^{6g-6+2r}$, where
$\cal U$ is an open set of the Teichm\"uller Space $ T^r_g \sim
B^{6g-6+2r}$. On the other hand, by patchworking of suspensions, we
will produce spacetimes with the same masses of certain hyperbolic
suspensions but which are not equivalent to any distinguished
deformation of them. In fact we will guess that they belong to
different connected components of the parameter space. So gravity
coupled to particles seems to be much more flexible than
pure gravity. In the last section we will state several related
questions and we will develop few speculations.
\medskip

\noindent Several constructions concerning Minkowskian suspensions
run, with minor changes, as in the matter-free case; so we will refer to
the diffuse paper {\bf [BG 1]} for more details.

\section {Geometric Surfaces with Conical Singularities.}

{\bf Cone points.} The local models of {\it flat} or {\it hyperbolic}
surfaces at a conical singularity are respectively given, in complex
coordinate, by the metrics on $\{|z|<1\}$ (set $\alpha > 0$):
$$ ds^2_{(E,\alpha)} = \alpha^2|z|^{2\alpha - 2}|dz|^2\ ,$$
$$ds^2_{(H,\alpha)} = 
\alpha^2[2/(1 - |z|^{2\alpha})]^2|z|^{2\alpha - 2}|dz|^2 \ .$$
\noindent They are obtained by pull-back of the standard Euclidean or 
Poincar\'e metrics on $\{|w| < 1\}$ via the map $w = z^{\alpha}$.

\noindent In both cases the  {\it concentrated curvature} at the 
conical point is $ k = 2\pi(1 - \alpha)$, the {\it cone angle} is 
$ 2\pi \alpha$. In order to have a genuine singularity, $\alpha \neq 1$. 
\medskip

\noindent {\bf Geometric cone surfaces.} It is convenient to adopt the 
formalism of geometric $(X,G)$-manifolds (see, for instance, chapter B
of {\bf [B-P]} or section 3 of {\bf [B-G 1]}).  Fix a {\it base}
compact oriented surface $F_g$ of genus $g\geq 0$ and fix
$p_1,\dots,p_r$ points on $F_g$. A {\it marked geometric (i.e. flat or
hyperbolic) surface with conical singularities}, of cone angles $2\pi
\alpha_i$, $i = 1,\dots,r$, is a homeomorphism
$$ \phi : (F_g,\{p_i\}) \to (S,\{q_i\}) $$
\noindent such that $S' = S\setminus \{q_i\}$ is a  $(\mr^2,Isom^+(\mr^2))$- 
(resp. a  $(\mh^2,Isom^+(\mh^2))$-) surface and its metric completion has
a conical singularity of cone angle $2\pi \alpha_i$ at $q_i$. 
\medskip

\noindent {\bf Gauss-Bonnet constraint.} 

\noindent The classical Gauss-Bonnet formula leads to the following
relations.

\noindent Flat Case:

\noindent ({\it Gauss-Bonnet equality})
$$ \sum_i k_i = 2\pi \sum_i(1-\alpha_i) = 2\pi (2 - 2g).$$

\noindent  Hyperbolic Case:  
$$ \sum_i k_i = 2\pi \sum_i(1-\alpha_i) = 2\pi (2 - 2g) + Area(S).$$

\noindent whence:

\noindent ({\it  Gauss-Bonnet inequality})
$$  \sum_i(1-\alpha_i) >  2 - 2g.$$

\noindent This implies, in any case, that when $g = 0$, necessarily
$r \geq 3$, and we will make this assumption by default.  
We say that 
$$\delta  = (\mx,g,[\alpha]_r) = (\mx,g,(\alpha_1,\dots, \alpha_r))$$
\noindent (where $\mx=\mr^2$ or $\mh^2$, $g\geq 0$ and the $\alpha_i$' 
satisfy the appropriate Gauss-Bonnet equality or inequality), is a 
{\it virtual type} of geometric surfaces with conical singularities. 
For a fixed type $\delta$ we denote by $T_{\delta}$ the 
{\it Teichm\"uller space} of marked surfaces of type $\delta$,
(regarded up to {\it Teichm\"uller equivalence} - see, for instance,
section 4 of {\bf [B-G 1]} for more details).  

\noindent When $r>0$, the fundamental group 
$\pi (F'_g = F_g\setminus \{p_i\})$ is a non Abelian free group with 
$s = 2g + r +1$ generators.
For each $[\phi]\in  T_{\delta}$ it is well defined, up to conjugation,
the {\it holonomy} representation
$$ \rho_{[\phi]}:\pi (F'_g)\to Isom^+(\mx). $$
\noindent The universal covering $p: S^*\to S$ is, in a natural way, 
a local isometry so that $S^*$ is homeomorphic to $\mr^2$ and it is 
endowed with a geometric structure with conical singularities.
We have also the {\it developing map} (well defined up to left action
of $Isom^+(\mx)$)
$$ D_{[\phi]}: (S')^*\to \mx\ .$$
\noindent $(S')^*$ is also homeomorphic to $\mr^2$ and is endowed with
a smooth geometric structure.
We can choose the representatives in such a way that,
for every $\gamma \in \pi (S')$, for every $x \in (S')^*$,
$$ D_{[\phi]}(\gamma (x)) =   \rho_{[\phi]}(\gamma)( D_{[\phi]}(x)). $$
\noindent  $D_{[\phi]}$ is a local isometry; when $r > 0$, $S'$ and $(S')^*$ 
are not (metrically) complete and, equivalently,  $D_{[\phi]}$ is not a global
isometry. 
\medskip

\noindent {\bf Orbifolds.} Geometric 2-dimensional compact orbifolds 
(with only conical singularities) make a special class of surfaces we
are concerned with.  Such an orbifold $S$ is a quotient $\mx/\Gamma$
where $\Gamma$ is a group of isometries of $\mx$ acting properly
discontinuosly and such that the set of points with non trivial
stabilizer is made by isolated points. For a genuine orbifold this set
is nonempty. They are classified as follows (see {\bf [T], [Sc]}).

\begin{prop} A geometric cone surface is a genuine Euclidean orbifold iff
it is of one of the types $(\mr^2,0,(1/2,1/3,1/6))$,
$(\mr^2,0,(1/2,1/4,1/4))$, $(\mr^2,0,(1/3,1/3,1/3))$,
$(\mr^2,0,(1/2,1/2,1/2,1/2))$.  A geometric cone surface is a genuine
hyperbolic orbifold iff it is of a type $(\mh^2,g,[\alpha]_r)$
satisfying the Gauss-Bonnet inequality and such that each $\alpha_i
\in [\alpha]_r$ is of the form $\alpha_i = 1/n_i$, $n_i \in
\mn^*$. Moreover all these types are actually realized by orbifolds.
\end{prop}
\medskip 

\noindent {\bf Conformal structures.} Associated to each geometric 
structure with conical singularities there is a natural {\it conformal
structure}: a conformal atlas is simply obtained by adding a chart in
complex coordinates as above at each conical singulatities, to any
atlas of the geometric structure on $S'$ (use the Poincar\'e disk
model for $\mh^2$).  So, for each virtual type $\delta$, if $T_g^r$ is
the classical {\it Teichm\"uller space} of conformal structures on
$F_g$, relatively to the marked points $p_i$ (which, as it is
well known, is homeomorphic to an open ball $B^{6g-6+2r}$), there is a
natural continuos map (in fact in the case of flat surfaces we take
off simple rescaling {\it by normalizing the area to be equal to
$1$})
$$ \psi_{\delta}: T_{\delta} \to T_g^r.$$

\noindent Geometric surfaces with conical singularities are classified 
by the following proposition.

\begin{prop} For any virtual type $\delta$, the natural map 
$ \psi_{\delta}$ is a homeomorphism.
\end{prop}      

\noindent The flat case is due to Troyanov (see {\bf [Tr]}). 
The orbifold case is treated in {\bf [T]}.
Let us sketch the main steps of a  proof in the general hyperbolic case.

{\bf (1)} {\it dim ($T_{\delta}) =$ dim ($T_g^r$).}

\noindent Let us outline first a way to construct all hyperbolic cone 
surfaces.  Fix $(F_g,\{p_1,\dots,p_r\})$ as before.  A {\it standard
spine} of $F'_g$ is a 1-complex $P$ embedded in $F'_g$, with only
3-valent vertices, such that $F'_g$ is a regular neighbourhood of $P$
($F'_g$ retracts onto $P$). Associated to such a $P$ there is a dual
(topological) {\it ideal triangulation} $\tau_P$ of $F_g'$, that is a
``relaxed'' (i.e. multiple and self adjacencies between triangles are
allowed) triangulation of $F_g$, having $\{p_1,\dots,p_r\}$ as
set of vertices. If $v(P) = |V(P)|$ denotes the number of vertices of
$P$ (i.e. the number of triangles of $\tau_P$), $e(P) = |E(P)|$ the
number of its edges (i.e. the number of the edges of the dual
triangulation), one has $ 3v(P) = 2e(P)$ so that $e(P) = 6g-6+3r$.
Clearly spines exist. Fix a spine $P$. For any {\it admissible} map
$f: E(P)\to \mr^+$ (i.e. a map such that at each vertex $v\in P$ the
values of $f$ on the three edges emanating from $v$ satisfy the
{\it triangular inequalities}), we can construct a hyperbolic surface with
$r$ conical singularities. This is obtained as a geometric realization
of the dual triangulation $\tau_P$, by using hyperbolic triangles with
edge lengths prescribed by $f$. Recall that each hyperbolic triangle
is determined by the edge lengths as well as by the interior angles,
and there are classical explicit formulas relating lengths and
angles. It is not too hard to see that varying the spine and the
admissible function, one can realize all the hyperbolic virtual
types. On the other hand, any cone hyperbolic surface arises in this
way. In fact let $(F_g,F_g')\sim (S,S')$ be such a surface. Consider
the subset $Q$ of $S$, such that for each $x\in Q$ there exist $i\neq
j$ such that $d(x,p_i) = d(x,p_j)$. Generically $Q$ is a standard
spine of $S'$; the interior of an edge of $Q$ consists of the points
with exactly two equidistant marked points $p_i,p_j$, the same along
the given edge. The ``axis'' of each edge, that is the geodesic arc
connecting $p_i$ and $p_j$ and passing from the point of the edge of
minimal distance from them, are the edges of a geometric realization
on the dual triangulation $\tau_Q$. In general $Q$ is a spine,
possibly with higher valency vertices; the same procedure produces a
dual ideal cellularization of $S'$ by convex hyperbolic polyhedra and
we eventually obtain a geometric triangulation by subdividing without
introducing new vertices. If a virtual type $\delta$ is realized by
an admissible map $f_0$ on $E(P)$, the maps realizing the same type are
obtained by imposing $r$ independent conditions. So one can deduce, at
least, that $T_{\delta}$ is a topological manifold of the right
dimension $6g - 6 + 2r$.
\medskip

{\bf (2)} {\it The map $ \psi_{\delta}$ is injective.}

\noindent Consider $\mh^2$ in the Poincar\'e disk model $D = \{|z|<1\}$,
and let $e^{2h}|dz|^2$ be the standard Poincar\'e distance. Realize 
a given element $\sigma$ of $T^r_g$ by a smooth hyperbolic surface (with marked
points) $S = D/\Gamma$. Two hyperbolic surfaces with conical singularities
of the same type, both representing $\sigma$, are given by two metrics
$e^{2(h+h_i)}|dz|^2$, $i=1,2$, such that each $h_i$ is a $\Gamma$-equivariant
function on $D$, with the same kind of singularities over the marked points
of $S$. It follows that $h_1 - h_2$ is a real analytic  $\Gamma$-equivariant
function on $D$ satisfying the Liouville equation 

$$ \Delta (h_1 - h_2) = e^{2h}(e^{2h_1} - e^{2h_2}).$$

\noindent As $S$ is compact $h_1 - h_2$ has maxima and minima. Either 
$ \Delta (h_1 - h_2) > 0$ near a maximum, or  $\Delta (h_1 - h_2)\leq 0$
near a minimum. By the maximun principle $h_1 - h_2$ is constant near the
minumum or the maximun and hence it is constant (and necessarily equal to $0$)
everywhere.
\medskip

{\bf (3)} {\it Conclusion.}  By the {\it invariance of domain}
theorem, $ \psi_{\delta}$ is a homeomorphism onto a non empty open subset of 
$T_g^r$.
To conclude it is enough to show that the image of $ \psi_{\delta}$ is closed.
This can be done by studing the convergence of the conformal factors (see
the above step), or by arguing (via geometric considerations) that the image
of a ``diverging'' sequence in $T_{\delta}$ is diverging in $T_g^r$.

\section {Minkowskian Suspensions.}

\noindent {\bf Particle world lines.} Let us give, first, the local 
models of the line of universe of a massive particle. They are
obtained by ``suspension'' of the local models for geometric cone
surfaces. We can take indifferently, in coordinates $(z,t)$,
$$ d\sigma^2_{(E,\alpha)} = - dt^2 + ds^2_{(E,\alpha)}\ ,$$
\noindent or, assuming  $t>0$
$$d\sigma^2_{(H,\alpha)} = - dt^2 + t^2 ds^2_{(H,\alpha)}\ .$$
\noindent They are equivalent {\it as local models}, in the sense that 
any point $(0,t_1)$ in the first model and any point $(0,t_2)$ in the
second one  have isometric neighbourhoods. They are not equivalent as 
global spacetimes; for instance if we take the time orientation in 
accordance with the $t$ coordinate, the canonical (CTF) of the first 
spacetime is
degenerate, constant equal to $\infty$, while $t$ is the (CTF) of the 
second one. We have a well defined cone angle $2\pi \alpha $ 
along such 
a universe line, which corresponds to a spacetime curvature concentrated 
along the line. In accordance with {\bf [D-J- `T H], [`T H]}, 
if we normalize
the gravitational constant to be $G=1$, the {\it mass} of the particle is 
related to the cone angle by $m = (1/4)(1-\alpha) $; 
in $(2+1)$-gravity there are not physical constraints on the sign of $Gm$, 
so that an arbitrarily big $\alpha$ is allowed. 
\medskip

\noindent {\bf Spacetimes with gravitating particles.} A marked globally 
hyperbolic spacetime (coupled to massive particles) of type
$$ \delta = (g,[\alpha]_r) = (g,(\alpha_1,\dots, \alpha_r))$$
\noindent is an homeomorphism
$$ \phi : (F_g\times \mr,\{p_i\}\times \mr) \to (M,L_i) $$
\noindent such that $M' = M\setminus \{L_i\}$ is an oriented and 
time-oriented globally hyperbolic flat Lorentzian 3-manifold (i.e. a
$(\mm^{2+1}, Isom^+(\mm^{2+1})$-manifold, where $\mm^{2+1}$ is the standard
Minkowski space) and each point of $L_i$ has a neighbourhood isometric
to the above local models, with cone angle $2\pi \alpha_i$. It is convenient
to restrinct to {\it Geroch marking}, that is we stipulate that the 
surfaces $\phi(F_g\times \{t\})$ are future Cauchy surfaces. 
As usually we work up to  Teichm\"uller equivalence and we denote by 
$T^{GR}_{\delta}$ the corresponding  Teichm\"uller space for a given type. 
To make it more meaningful it is convenient to restrict to maximal 
spacetimes. Identifying $F_g$ with $F_g \times \{0\}$ we have also the 
holonomy representation
$$ \rho_{[\phi]}:\pi (F'_g)\to Isom^+(\mm^{2+1}). $$
\noindent We also make the current assumption that the linear part 
of the holonomy takes values in $SO^+(2,1)$, the group of Lorentz
transformations keeping the upper half-space invariant.
\medskip

\noindent {\bf  Minkowskian suspensions of geometric cone surfaces.} They are
peculiar spacetimes which actually are $(Y,G(Y))$-manifolds, for
suitably chosen open subsets $Y$ of $\mm^{2+1}$, $G(Y)$ being the
group of Minkowskian isometries keeping $Y$ invariant. As $Y$ we will
take:
$$Y_E = \mm^{2+1}$$
\noindent with metric $(dx^1)^2 + (dx^2)^2 -(dx^3)^2$, and thought fibred by 
the planes $\{x^3 = a\}$.  
$$Y_H = \{x \in \mm^{2+1}:\  (x^1)^2 + (x^2)^2 - (x^3)^2 < 0,\ x^3>0 \}$$
\noindent thought fibred by the surfaces 
$$ \{x \in \mm^{2+1}:\ (x^1)^2 + (x^2)^2 - (x^3)^2 = -a^2,\ x^3>0\}$$

\noindent finally
$$Y_T = \{x \in \mm^{2+1}:\ (x^1)^2  - (x^3)^2 < 0,\ x^3>0 \}$$
\noindent thought fibred by the surfaces 
$$ \{x \in \mm^{2+1}:\ (x^1)^2  - (x^3)^2 = -a^2,\ x^3>0\}.$$
\noindent By the change of coordinates
$$ x^1 = \tau sh(u),\ x^2 = y,\ x^3 = \tau ch(u) $$
\noindent we see that $Y_T$ is isometric to
$$ P = \{ (u,y,\tau)\in \mr^{2+1}: \tau >0\},\ {\rm with\  metric} 
\ \tau^2du^2 + dy^2  - d\tau^2 $$
\noindent and $P$ is fibred by the level planes of $\tau$.

\noindent Each $Y_*$ is  oriented and time-oriented in the usual way.

\noindent If $S$ is a flat cone surface of type $(\mr^2,g,[\alpha]_r)$,
its Minkowskian suspension $M(S)$ is the obviously associated 
$(Y_E,G(Y_E))$-spacetime of type $(g,[\alpha]_r)$, with holonomy equal to 
the holonomy of $S$. It is fibred by parallel copies of $S$. 
The canonical (CTF) is degenerate, costant equal to $\infty$.
These are called {\it static Minkowskian suspensions}.
\medskip

\noindent If $S$ is a hyperbolic cone surface of type  
$(\mh^2,g,[\alpha]_r)$, its Minkowskian suspension  $M(S)$ is the associated 
$(Y_H,G(Y_H))$-spacetime of type $(g,,[\alpha]_r)$, with
holonomy equal to the holonomy of $S$. It is fibred by parallel rescaled 
copies of $S$; these surfaces are the level surfaces $S_a$ ($S = S_1$)
of the canonical 
(CTF); out of the particles they have constant mean curvature $1/a$ and 
constant intrinsic curvature equal to $-1/a^2$. The initial singularity 
consists of one point.
\medskip

\noindent These suspensions are particularly nice when $S$ is an orbifold 
(and the matter-free spacetimes are particular cases); if the orbifold 
$S = \mx/\Gamma$,
$\Gamma$ acts isometrically also on the corresponding $Y_*$, and 
$M(S) = Y_*/\Gamma$. The parameter space of $Y_E$ or $Y_H$-suspensions 
of a given type coincides, tautologically, with the parameter space of the 
suspended geometric cone surfaces (see the previous section).
\medskip
 
\noindent The $Y_T$-Minkowskian suspensions involve the special flat cone 
surfaces given by the {\it meromorphic quadratic differentials} with
at most  simple poles on Riemann surfaces. In fact each such a
suspension is determined by a couple $(F,q)$, where $F$ is a Riemann
surface and $q$ is a quadratic differential. That is, it is determined
not only by the cone surface, but also by the horizontal and vertical
measured foliations of the quadratic differential. We have already
studied such spacetimes in {\bf [B-G 2]} where we have shown how they
``materialize'' the classical {\it Teichm\"uller flow}. See also {\bf
[B-G 1]} for a description of the (CTF). In fact in {\bf [B-G 2]} we
considered only holomorphic quadratic differentials, but everything
runs verbatim if one allows also simple poles. Recall that in this way
one can realize all the types with $2\pi \alpha_i = n_i\pi,\ n_i \geq
1$, satisfying the Gauss-Bonnet equality, with  four exceptions
(see {\bf [M-S]}).  Moreover, for any given realizable type, one knows
the degrees of freedom (see {\bf [V]}): if $\mu (a)$ denotes the
number of cone points of cone angle $a$, then the degrees of freedom
are
$$ 2g + \sum \mu(a) + (\epsilon - 3)/2$$

\noindent where $\epsilon = -1$ iff there is at least one cone angle
with odd $n_i$, and it is equal to $1$ otherwise. For example, when
the type contains only $n_i = 3$ (this corresponds to holomorphic 
quadratic differentials with simple zeros), the dimension of the 
corresponding space of $Y_T$-suspensions is $6g-6$.    

\noindent The only orbifolds which produce such a kind
of suspension are the orbifolds of type $(\mr^2,0,(1/2,1/2,1/2,1/2))$.
They are obtained by the natural identification of the edges of 
two copies of a same Euclidean rectangle. The corresponding group $\Gamma$ 
is generated by two orthogonal translations and the rotation of angle $\pi$. 
Groups that determine the same $Y_E$-suspension (up to equivalence), 
do determine in general different $Y_T$-suspensions; in fact if we look at 
these groups acting on $P$, the horizontal and vertical foliations on each 
$\tau$-level plane induce different foliations on the (CTF)-level 
surfaces of the two suspensions.

\section {Distinguished Deformations of Hyperbolic Suspensions.}

\noindent In this section we will refer heavily to {\bf [B-G 1]} and 
to {\bf [Me]}.  All matter-free spacetimes with space of genus $g\geq
2$ are obtained by specific ``deformations'' of Minkowskian
suspensions $M(S)=Y_H/\Gamma$, and each such a deformation $M(S,{\cal
F})$ is uniquely determined by a {\it measured geodesic lamination}
$\cal F$ on $S$. $M(S)$ and $M(S,{\cal F})$ have holonomies with the
same linear part.  The lifted lamination ${\cal F}^*$ to the universal
covering $S^* = \mh^2$ is ``dual'' to a {\it real tree} which is
isometric to the initial singularity of $M(S,{\cal F})^*$. The initial
singularity of $M(S,{\cal F})$ must be properly interpreted in terms
of a natural action of the fundamental group $\pi(S)$ on this real
tree; the natural actions of $\pi(S)$ on the universal covering of the 
level surfaces of the (CTF) of  $M(S,{\cal F})$, asymptotically degenerate
to that action on the real tree.

\noindent  If $S$ is a hyperbolic cone surface of type 
$\delta = (g,[\alpha]_r)$ (we have omitted the ``$\mh^2$'' in $\delta$), 
and $\cal F$ is a measured geodesic lamination with {\it compact support} 
in $S'$ we can repeat those constructions (working with $S^*$ which is now a 
cone hyperbolic surface) getting, by definition, a 
{\it distinguished deformation} $M(S,{\cal F})$ of $M(S)$, which is again 
a spacetime of type $\delta$. 

\noindent The simplest deformations arise when $\cal F$ is a 
{\it multicurve}, i.e. it consists of the finite union of disjoint simple 
geodesics endowed with positive weights. Assume, for simplicity, that 
there is one geodesic $\sigma$, with weight $s$ and length $r$. 
Consider the quotient $A'(s,r)$ of 
$B'(s,r)=\{(u,y,\tau)\in P;\ 0\leq y \leq s\}$ by the group generated by the 
translation $(u,y,\tau)\to (u+r,y,\tau)$. Actually it is better to consider 
the isometric quotient $A(s,r)$ of $B(s,r)\subset Y_T$, obtained via the 
explicit change of coordinates given in section 3.  Then, to construct 
$M(S,{\cal F})$, cut-open $M(S)$ along the suspension of $\sigma$ and 
insert $A(s,r)$ in the natural way. $M(S,{\cal F})$ is, by construction, 
fibred by $C^1$-embedded space-like surfaces (made by the union of pieces 
of constant negative curvature and flat annuli); in fact they are the level 
surfaces of the canonical (CTF).

\noindent The above construction is very simple nevertheless, as multicurves 
are dense in the space of measured geodesic laminations, by
making the multicurve ``complicated'' enough, we can fairly well approximate 
the shape of any general distinguished deformation.
\medskip
 
\noindent Given a hyperbolic type $\delta = (g,[\alpha]_r)$, we 
denote by $D(\delta)$  the subset of $T^{GR}_{\delta}$ determined by the
distinguished deformations of Minkowskian suspensions of hyperbolic cone
surface of type $\delta$. Of course, a suspension is meant as the trivial 
deformation of itself and there is a natural projection 
$p: D(\delta)\to T_{\delta}$. 
The following proposition gives partial information
on $D(\delta)$. We will use some notations introduced in section 2. The set 
of hyperbolic ``$(g,r)$-types'' can be identified with an open
set of $\mr^r$. 

\begin{prop}
{\rm (1)} For each hyperbolic type $\delta$ there is an open (possibly
empty) maximal subset ${\cal U}_{\delta}$ of $T_{\delta}$ such that
$p^{-1}({\cal U}_{\delta})\subset D(\delta)$ is homeomorphic 
to ${\cal U}_{\delta} \times \mr^{6g-6+2r}$ (and $p$ becomes the natural 
projection onto the first factor).  
\medskip

\noindent {\rm (2)} For each $(g,r)$ there is a maximal non empty open subset 
${\cal W}_{(g,r)}$ of the space of $(g,r)$-types, such that for each 
$\delta \in {\cal W}_{(g,r)} $, ${\cal U}_{\delta}$ is non empty. 
\medskip

\noindent {\rm (3)} For any $\delta$, 
$$12g-12+4r \geq dim D(\delta) \geq 6g-6+2r.$$

\noindent {\rm (4)} If  ${\cal U}_{\delta}$ is non empty, 
then ${\cal U}_{\delta} \times \mr^{6g-6+2r}$ is an open subset of  
$T^{GR}_{\delta}$.
\end{prop}

\noindent Let us give a scketch of a proof.
By using the result of section 2, the first statement is equivalent to
show that the space of measured geodesic laminations with compact
support on $S'$, for a given hyperbolic cone surface $S$ in $\cal U$
(for a suitable $\cal U$), is homeomorphic to $\mr^{6g - 6 +
2r}$. This fact is known in the ``limit'' case when each $\alpha_i =
0$, that is when $S'$ is a complete finite area hyperbolic surface
with $r$ cusps (see {\bf [Pe]}). Let us denote by $HT^r_g$ (which is
homeomorphic to $T^r_g$ ) the Teichm\"uller space of such hyperbolic
surfaces with $r$ cusps and fix one surface $F$. It is known that each
geodesic lamination with compact support on $F$ has support contained
in $F''$ obtained by removing from $F$ all the horocycles of length
$<1$ around all the cusp points (see {\bf [Pe]} pag. 72). It turns out
that any hyperbolic cone surface $S$ which is ``geometrically'' close
to $F$ has, up to homeomorphism, the same space of measured geodesic
laminations with compact support on $S'$.  The crucial fact is that if
$S$ is close enough to a cusped $F$, each isotopy class of essential
(i.e. non contractible nor contractible to one cone point) simple
curves on $S'$ has a simple geodesic (in $S'$) shortest length
representative.  For some notions on the ``geometric topology'' see,
for instance, chapter E of {\bf [B-P]}. In our situation ``$S$
geometrically close to $F$'' roughly means that, removing suitable
``round'' disks with centres at the cone points of $S$, we find $S''$
which is {\it quasi-isometric} to $F''$, by a quasi-isometry close to
an isometry.  It follows that for any fixed compact subset $K$ of
$HT^r_g$ there is an open subset $U_K$ (possibly empty) of
$T_{\delta}$, which satisfies the first statement of the proposition.

\noindent To prove the second statement, it is enough to show that,
for any fixed $F$ as before, there are cone surface $S$ close to $F$
in the above sense. Fix a geodesic ideal triangulation $\cal T$ of $F$
(i.e.  a ``relaxed'' triangulation of $F$ by ideal hyperbolic
triangles).  For each $0<a<1$ consider the horocycles of length $a$
around the cusps of $F$.  Associate to each edge of the triangulation
the length of the subarc determined by the horocycles. Consider the
cone surface $S$ obtained accordingly with the construction after 
proposition 2.2, by using the same $\cal T$ as topological ideal 
triangulation of $F_g'$ and those lengths as edge-lengths. If $a$ is small 
enough, $S$ is close to $F$.
 
\noindent 
$S$ is not close enough to a cusped $F$ when, at a very qualitative
level, the masses are not big enough or the particles are too close
each other on a given level surface of the (CTF) of the corresponding
Minkowskian suspension. In such a case the basic trouble consists in
the fact that the shortest length representative (if any) of an essential
isotopy class of simple curves on $S'$ might be a broken geodesic
passing through some cone points or not even a simple curve.

\noindent The third statement is clear from the above discussion.

\noindent To achieve the last  statement it is enough to show that 
$T^{GR}_{\delta}$ is of dimension $12g - 12 + 4r$; we are going to
argue it without any assumption on the spacetime type $\delta=
(g,[\alpha]_r)$.
\medskip

\noindent {\bf The degrees of freedom of $T^{GR}_{\delta}$.} Fix a marked 
spacetime $M$ of type $\delta$ and a relatively compact globally
hyperbolic open neigbourhood $U$ of the Cauchy surface image of
$F_g\times \{0\}$.  Let $\rho: \pi(F_g')\to ISO^+(2,1)$ be its
holonomy. As $\pi(F_g')$ is a free group, a deformation of
$\rho$ is simply obtained by modifying $\rho$ on a set of $2g-2+r+1$
free generators. If a deformation $\rho'$ is small enough, then, by
the stability property of holonomies, $\rho'$ is still the holonomy of a
spacetime structure on the interior of $U$, with $r$ gravitating
particles. So, as the holonomy is defined only up to conjugation, the
dimension of the set of all these spacetimes ``close'' to $M$ is
$12g-12+6r$. In order to impose that the spacetimes have the specific
cone angles prescribed by $\delta$, we have to impose $2r$ (that is
$(6 - d)r$, where $d$ is the dimension of the conjugation orbit of a
``rotation'') more independent conditions, and we finally get the
required number of degrees of freedom $12g-12+4r$.
\medskip

\section { Patchworking of Minkowskian Suspensions.}

\noindent A simple variation of the costruction of the distinguished
modification of hyperbolic suspensions, based on multicurves, 
that we have described in the previous section, will produce interesting 
new examples of spacetimes. 

\noindent Let $M(S,{\cal F})$ be as in the previous section. 
Assume that we have a finite union of simple closed geodesic on $S'$
{\it disjoint from ${\cal F}$}. For simplicity, assume that there is a
single geodesic $\sigma$ of length $a$. Let $(F,q)$ be a Riemann
surface with a meromorphic quadratic differential $q$, with at most
simple poles.  Let $M(F,q)$ be the corresponding $Y_T$- Minkowskian
suspension (see section 3). Assume that the q-horizontal foliation on
$F$ contains a simple closed leaf $c$ of length $a$. Then we can
construct new spacetimes as follows: cut-open $M(S,{\cal F})$ along
the suspension of $\sigma$ and $M(F,q)$ along the suspension of $c$;
then glue, pairwise, pieces of $M(S,{\cal F})$ with pieces of
$M(F,q)$ along isometric boundary components in the natural way.
Note that there is, in general, a finite number of possible
combinations, and the resulting Lorentz manifolds may be not
connected, so we can take each connected component as a new
spacetime. Call $M([S,{\cal F},\sigma],[F,q,c])$ any spacetime
obtained in this way.  By construction, it is fibred by space-like
surfaces (made by rescaled pieces of $S$ and by ``stretched '' pieces
of $F$) which actually are the level surfaces of the canonical (CTF)
of $M([S,{\cal F},\sigma],[F,q,c])$. Note also that the construction
can be {\it iterated}, starting from suitable $M([S,{\cal
F},\sigma],[F,q,c])$; so one can produce a wide class of new
examples. This {\it patchworking} is peculiar of spacetimes with
gravitating particles; in fact if we formally apply it to matter-free
spacetimes we get nothing else than distinguished deformations of
hyperbolic suspensions.

\noindent  In particular, let us use as $(F,q)$ the orbifolds of type  
$$(\mr^2,0,4[1/2]) = (\mr^2,0,(1/2,1/2,1/2,1/2))$$
\noindent  with the horizontal and
vertical foliations of $q$ (with $4$ simple poles) parallel to the
edges of the ``fundamental'' rectangle. It is not hard to construct by the
patchworking procedure hyperbolic types of the form 
$\delta = (g,[\alpha]_r) = (g, [\alpha]_{r'}\cup 2h[1/2])$, $2h+r'=r$. 
On the other hand, these new spacetimes do not belong to
$D(\delta)$ because, for instance, the level surfaces af the canonical
(CTF) are not isometric (there are cone points of cone angle $\pi$
with no isometric neighbourhoods). Other differences manifest
themselves by studying the past asymptotic states of the respective
(CTF). By small perturbation of the holonomy of these examples one
could produce examples out of $D(\delta')$ for any $[\alpha]'_r$ close
to $[\alpha]_r$.

\section {Final Questions and Considerations.}  

\noindent We are going to conclude with some questions, problems  and, 
sometimes, with a guess about them.
\medskip

{\it {\rm (1)} Is $T^{GR}_{\delta}$ connected ?}

\noindent The answer could depend on the type. We guess that the above
examples not belonging to $D(\delta)$, actually do not even belong 
to the same connected component of any element of $D(\delta)$.
\medskip

{\it {\rm (2)} Does any spacetime satisfy the Gauss-Bonnet constraint
$ \sum (1-\alpha_i) \geq 2-2g$ ?}

\noindent We guess that by suitable small perturbations of the holonomy of 
static Minkowskian suspensions (which satisfy the Gauss-Bonnet 
{\it equality}) one could obtain spacetimes with 
$\sum_i (1-\alpha_i) < 2-2g$ .    
\medskip

\noindent We note that all the examples of spacetime that we have 
produced starting from non static Minkowskian suspensions have the
following property : 

{\it each particle line of universe has a neighboourhood isometric to
the set of points of {\rm spatial distance} $<bt$, for some positive
$b$, from the $t$-axis in the model $\{(z,t),\ |z|<1,\ t>0\}$ with
metric $d\sigma^2_{(H,\alpha)}$ (see section 3).}

{\it {\rm (3)} Does the same property hold for any spacetime with {\rm
tame} - see {\bf [B-G 1]}- canonical {\rm (CTF)} with values onto
$(0,\infty)$?}

\noindent It would be interesting to find, if any, examples where the linear
function $bt$ must be replaced by some positive function $f(t)$ going 
faster to $0$ when $t\to 0$.  
\medskip

{\it {\rm (4)} Find an intrinsic characterization of hyperbolic cone
surfaces belonging to ${\cal U}_{\delta}$.}

\noindent One expects that it could be expressed in terms of inequalities
involving the cone angle, the genus and the distances between the cone
points.
\medskip

{\it {\rm (5)} Describe ${\cal W}_{(g,r)}$. In particular, does
$m(g,r)$ exist, with $1>m(g,r)>0 $, such that for any $\delta \in {\cal
W}_{(g,r)} $ and for any mass $m_i$ associated to $\delta$, one has
$m_i > m(g,r)$ ?}

\noindent For example, beside the ``rigid'' case  $(g,r) = (0,3)$,
the very peculiar case $(g,r) = (0,4)$ has ${\cal W}_{(0,4)}$ which
coincides with the whole space of $(0,4)$-types; moreover for each
type $\delta \in {\cal W}_{(0,4)}$, ${\cal U}_{\delta}$ coincides with
$T_{\delta}$, so that $D(\delta)$ = $T_{\delta} \times
\mr^2$. On the other hand, we guess, for example, that for
each $(0,r)$, $r>4$, the last question has negative answer.
\medskip

{\it {\rm (6)} Is $D(\delta)$ always of dimension $> 6g - 6 +2r$ ?}

\noindent In other words, one is asking if there are always non trivial
distinguished deformations. We guess that when $g\geq 2$ and the
masses are all positive, then $D(\delta)$ contains at least
$T_{\delta}\times \mr^{6g-6}$; in other words one expects that
there is at least the same ``amount'' of distinguished deformations of
the matter-free case of the same genus.
\medskip

{\it {\rm (7)} Let $C$ be any closed subset of ${\cal U}_{\delta}$. Is
$p^{-1}(C)\subset  {\cal U}_{\delta}\times  \mr^{6g-6+2r}$ closed in
$T^{GR}_{\delta}$ ?}   
\medskip

\noindent Finally we note that in several instances of the present paper 
we have seen how very natural perturbations of a given spacetime {\it
do not preserve the type} (see for instance the constructions of
section 2 or the argument at the end of section 4). It would suggest
that the study of (2+1)-gravity (coupled to particles) ``type by
type'', or even ``space-genus by space-genus'', could be misleading.
Spacetimes would be considered ``all toghether'' and it becomes quite
demanding to figure out the structure of the corresponding (infinite
dimensional) parameter space. We guess that Grothendiek theory of
``Teichm\"uller Towers'' could play an important role.

$${\bf REFERENCES}$$
\bigskip

\noindent {\bf[A-M-T]} L. Anderson - V.Moncrief - A. Tromba,
Journal of Geometry and Physics 23 (1997) 191-205.
\medskip

\noindent {\bf[B-C-V]}, A. Bellini - M. Cianfaloni - P. Valtancoli, 
Physics Lett. B 357 (1995) 532; Nucl. Phys. B 462 (1996) 453.
\medskip

\noindent {\bf[B-G 1]} R. Benedetti - E. Guadagnini, 
preprint gr-qc/0003055 .
\medskip

\noindent {\bf[B-G 2]} R. Benedetti - E. Guadagnini, Physics Letters B 441
(1998) 60-68.
\medskip

\noindent {\bf[B-P]} R. Benedetti - C. Petronio, {\it Lectures on 
Hyperbolic Geometry}, Springer-Verlag 1992.
\medskip

\noindent {\bf [D-J-'t H]}  S. Deser -  R. Jackiw - G. 't Hooft,
Ann.Phys. (NY) 152 (1984) 220.
\medskip

\noindent {\bf['t H]}, G. 't Hooft, Class. Quantum Grav. 10 (1993),
S79-S91.
\medskip

\noindent {\bf[Me]} G. Mess, Preprint IHES/M/90/28 (1990).
\medskip

\noindent {\bf[M-S]} H. Masur - J. Smillie, Comment. Math. Helvetici 68
(1993) 289-307.
\medskip

\noindent {\bf[Mn-S]}, P. Menotti - D. Seminara, preprint hep-th/9907111.
\medskip

\noindent {\bf[Mo]} V. Moncrief, J.Math.Phys. 30 (1989) 2907-2914.
\medskip

\noindent {\bf[Pe]}, R.C. Penner with J.L. Harer {\it Combinatorics
of Train Tracks}, Annals of Math. Studies 125, Princeton 1992.
\medskip

\noindent {\bf[Sc]} P. Scott, Bull. London Math. Soc. 15 (1983),no 5,
401-487.
\medskip

\noindent {\bf[T]} W.P. Thurston, {\it Geometry and Topology of 
3-manifolds}, ``The'' Notes, Princeton, 1982.
\medskip

\noindent {\bf[Tr]} M. Troyanov, L'Enseignement Math. t. 32 (1986), 79-94.
\medskip

\noindent {\bf[V]} W. A. Veech, Annals of Math. 124 (1986), 441-530.
\medskip

\noindent {\bf[W]} E. Witten, Nucl. Phys. {\bf B311} (1988) 46.
\medskip

\vspace{1cm}

\hspace{8.5cm} Dipartimento di Matematica

\hspace{8.5cm} Universit\`a di Pisa

\hspace{8.5cm} Via F. Buonarroti, 2

\hspace{8.5cm} I-56127 PISA (Italia)

\hspace{8.5cm} benedett@dm.unipi.it

\vspace{1cm}

\hspace{8.5cm} Dipartimento di Fisica

\hspace{8.5cm} Universit\`a di Pisa

\hspace{8.5cm} Via F. Buonarroti, 2

\hspace{8.5cm} I-56127 PISA (Italia)

\hspace{8.5cm} guadagni@difi.unipi.it

\end{document}